\documentclass[11pt,a4paper]{article}

\usepackage{amsmath,amsthm,amssymb}
\usepackage{graphics,graphicx}
\usepackage{epsfig}
\usepackage{multicol}
\usepackage{color}
\usepackage{a4wide}

\begin{document}

\rightline{PUPT-2252}
\rightline{QMUL-PH-07-20}
\vspace{2truecm}

\centerline{\LARGE \bf Goldstone Bosons and Global Strings }

\vspace{0.5cm}

\centerline{\LARGE \bf in a Warped Resolved Conifold}

\vspace{1.3truecm}

\centerline{
    {\large \bf I. R. Klebanov${}^{a,b}$},
    {\large \bf A. Murugan${}^{a}$},
    {\large \bf D. Rodr\'{\i}guez-G\'omez${}^{a,c}$}
    {\bf and}
    {\large \bf J. Ward${}^{d}$}}

\vspace{.4cm}
\centerline{{\it ${}^a$Department of Physics, 
Princeton University}}
\centerline{{\it Princeton, NJ 08544, USA}}

\vspace{.4cm}
\centerline{{\it ${}^b$Center for Theoretical Physics, 
Princeton University}}
\centerline{{\it Princeton, NJ 08544, USA}}

\vspace{.4cm}
\centerline{{\it ${}^c$ Center for Research in String Theory, Queen Mary University of
    London}} \centerline{{\it Mile End Road, London, E1 4NS, UK}}

\vspace{.4cm}
\centerline{{\it ${}^d$ Department of Physics and Astronomy, University of
    Victoria}} \centerline{{\it Victoria, BC, V8P 1A1, Canada}}

\vspace{2truecm}

\centerline{\bf ABSTRACT}
\vspace{.5truecm}

\noindent
A warped resolved conifold background of type IIB theory, constructed in hep-th/0701064, 
is dual to the supersymmetric
$SU(N)\times SU(N)$ gauge theory with a vacuum expectation value (VEV)
for one of the bifundamental chiral superfields. 
This VEV breaks both the superconformal invariance and the 
baryonic symmetry. The absolute value of the VEV controls the resolution
parameter of the conifold. 
In this paper we study the phase of the VEV, which corresponds to the Goldstone boson of the 
broken symmetry. We explicitly construct the linearized perturbation of the 
4-form R-R potential that contains the Goldstone boson. 
On general grounds, the theory should contain global strings which create a monodromy of 
the pseudoscalar Goldstone boson field. We identify these strings with the $D3$-branes wrapping the
two-cycle at the tip of the warped resolved conifold.

\newpage

\section{Introduction and summary}

The AdS/CFT duality \cite{Maldacena:1997re,Gubser:1998bc,Witten:1998qj} has produced 
major progress in our understanding of the intimate relationship between the dynamics of 
gauge theories and strings. 
The basic version of the duality is motivated 
by considering a stack of many $D3$-branes in flat space. 
The dual descriptions of $D3$-branes as a supergravity background solution on which strings propagate, or as objects carrying a worldvolume gauge theory, allows one to identify 
type IIB strings on $AdS_5\times S^5$ 
with $\mathcal{N}=4$ SYM theory. Furthermore 
one can engineer more elaborate (and less symmetric) versions of the duality
by considering $D3$-branes at the tip of a generic $CY$ cone. 
Among the simplest examples is the cone over $T^{1,1}$, the conifold, 
which leads to a duality between type IIB
strings on $AdS_5\times T^{1,1}$ with $N$ units of R-R 5-form flux
and a $SU(N)\times SU(N)$ SCFT coupled to bifundamental
chiral superfields $A_1, A_2, B_1, B_2$ \cite{Klebanov:1998hh}.

The singularity of the conifold can be smoothed out in two different ways: 
either by blowing up a 3-cycle, leading to the deformed conifold, 
or by blowing up a 2-cycle, leading to the resolved conifold. 
The deformation of the conifold may be produced by turning on $M$ units of R-R flux through the 3-cycle
\cite{KS,KT} (for reviews, see \cite{Herzog:2002ih}). 
The resulting warped product of $R^{3,1}$ and the deformed conifold gives 
rise to interesting phenomena arising from the fact that the corresponding 
$SU(N+M)\times SU(N)$ gauge theory undergoes a duality cascade
leading to confinement and chiral symmetry breaking in the infrared. 
The resolution of the conifold has a simpler interpretation: 
it corresponds to giving expectation values to the fields $A_i, B_j$ in the 
$SU(N)\times SU(N)$ gauge theory \cite{Klebanov:1999tb}. Recently a particularly simple
example where only one of the fields acquires a diagonal VEV, $B_2= u I_{N\times N}$, was
worked out in detail \cite{Klebanov:2007us}. The dual warped 
product of $R^{3,1}$ and the resolved conifold corresponds to a stack
of $N$ $D3$-branes placed at a point on the blown-up 2-sphere at the tip. 
The warp factor produced by the $D3$-branes was explicitly solved for. 

The VEV of $B_2$ spontaneously breaks the $U(1)_B$ symmetry of the theory. 
The gauge invariant order parameter for this breaking is the baryonic operator
${\rm det} B_2$. In \cite{Klebanov:2007us} the modulus of the VEV of this operator was computed 
holographically using a Euclidean $D3$-brane wrapping a holomorphic 4-chain. In this paper we will
discuss the phase of this VEV and phenomena associated with its variation.
One of these phenomena is the presence of a Goldstone boson; we will exhibit the normalizable
4-form perturbation around the warped resolved conifold that contains it.
The required perturbation contains a product of two 2-forms, an antisymmetric tensor in $R^{3,1}$
and a closed form $W$ on the warped resolved conifold. The dual of the antisymmetric tensor defines
the pseudoscalar Goldstone boson $p$. 
We will write down the equations that determine the 2-form $W$ and show that
they come from the minimization of a 
positive definite functional which, in the warped case and with appropriate boundary conditions, remains finite. 
In fact, this functional is the norm of the form $W$. 
The asymptotically
AdS warp factor is crucial for
the normalizability of $W$; in the unwarped case, the corresponding $W$ is not normalizable.

On general grounds, a broken global symmetry may give rise
to ``global'' strings associated with a non-trivial monodromy of the Goldstone 
boson.
We will show that, on the string side of the duality,
they are realized as $D3$-branes which wrap the finite $S^2$ at the tip of the 
warped resolved conifold, with the remaining two world volume directions lying within
$R^{3,1}$. Interestingly the tension of these 
strings is not sensitive to the warp factor. We will show that they are BPS saturated and are stable at 
the tip, $r=0$. Such wrapped $D3$-branes couple to the components of R-R 4-form described above, and
create the monodromy of the pseudoscalar $p$. We will also show that the phase of the baryonic
condensate computed holographically from $e^{-S_{E3}}$ is determined by $p$.

The outline of the paper is as follows. In section 2 we review the duality between the warped resolved
conifold and the gauge theory which was elucidated in \cite{Klebanov:2007us}. 
In section 3 we study global strings obtained from partially wrapped $D3$-branes and check their
$\kappa$-symmetry. In section 4 we derive the equations for the fluctuation of the R-R 
4-form potential containing the
Goldstone mode of the broken $U(1)_B$ symmetry, and solve them in various limits.
Section 5 is devoted to a further
analysis of spontaneous breaking of $U(1)_B$ using the dual supergravity. 
In particular, we calculate the Goldstone boson ``decay constant.''
We conclude with a few comments in section 6.

\section{The Warped Resolved Conifold} 

It is well-known that the conifold can be described by the equation
\begin{equation}
z_1^2+z_2^2+z_3^2+z_4^2=0\ ,\quad z_i\in \mathbb{C}\ .
\end{equation}
By defining a related set of $w_i$ coordinates, we can write this equation as
\begin{equation}
Z=\left(\begin{array}{c c} z_3+iz_4 & z_1-iz_4 \\ z_1+iz_4 & -z_2+iz_4\end{array}\right)=
\left(\begin{array}{c c} w_1 & w_3\\ w_4 & w_2\end{array}\right);
\ {\rm{det}}\, Z=0\ .
\end{equation}
It is possible to solve this by choosing
\begin{equation}
Z=\left(\begin{array}{c c} a_1b_1 & a_1b_2\\ a_2b_1 & a_2b_2\end{array}\right)
\end{equation}
and use the $(a_i,b_i)$ coordinates to describe the conifold. 
However they are not uniquely determined, since they are identified under
\begin{equation}
(a_i,b_i)\rightarrow (\lambda a_i,\frac{1}{\lambda}b_i)\ ,\quad \lambda\in C^*\ .
\end{equation}
In order to partially fix this freedom, one can impose 
the phase identification $a_i\sim e^{i\alpha}a_i$ and $b_i\sim e^{-i\alpha}b_i$. 
In the dual gauge theory this corresponds to the $U(1)_B$ symmetry which assigns 
opposite charges to $a_i$ and $b_i$. 
We are still left with the modulus of the above transformation, which we can fix by demanding 
$|b_1|^2+|b_2|^2-|a_1|^2-|a_2|^2=0$. In the $U(1)$ gauge theory
this constraint arises from the D-term, but as emphasized in \cite{Klebanov:1999tb}, 
in the $SU(N)\times SU(N)$ case the analogous constraint is absent.

A simple way to understand the resolution of the conifold is to deform the modulus constraint above into
\begin{equation}
\label{resolution}
|b_1|^2+|b_2|^2-|a_1|^2-|a_2|^2=u^2\ ,
\end{equation}
where $u$ is a real parameter which controls the resolution. The resolution corresponds 
to a blow up of the $S^2$ at the bottom of the conifold. 
In the dual gauge theory turning on $u$ corresponds to a particular choice of vacuum. 
After promoting the $a,b$ fields into the bifundamental chiral superfields
of the dual gauge theory, we can define the operator $\mathcal{U}$ as
\begin{equation}
\mathcal{U}=\frac{1}{N}{\rm{Tr}} (B_1^{\dagger}B_1+B_2^{\dagger}B_2-A_1^{\dagger}A_1-A_2^{\dagger}A_2)\ .
\end{equation}
Thus, the warped singular conifolds correspond to gauge theory vacua where 
$\langle \mathcal{U}\rangle=0$, while the warped
resolved conifolds correspond to vacua where $\langle \mathcal{U}\rangle\ne 0$. 
In the latter case, some VEVs for the bi-fundamental fields 
$A_i,B_j$ must be present. Since these 
fields are charged under the $U(1)_B$ symmetry, 
the warped resolved conifolds  
correspond to vacua where this symmetry is broken \cite{Klebanov:1999tb}.

A particularly simple choice is to give a diagonal VEV to only one of the scalar fields, say, $B_2$. 
As seen in  \cite{Klebanov:2007us}, this choice breaks the $SU(2)\times SU(2)\times U(1)_B$ 
symmetry of the CFT down to $SU(2)\times U(1)\times U(1)$. The string dual is given by a warped resolved
conifold 

\begin{equation}
ds^2=h^{-1/2}dx_{1,3}^3+h^{1/2}ds_6^2\ .
\end{equation}
The explicit form of the Calabi-Yau metric of the resolved conifold is given by \cite{Pando:2000sq}
\begin{eqnarray}
ds_6^2&=&K^{-1}dr^2+\frac{1}{9}Kr^2\Big(d\psi^2+\cos\theta_1 d\phi_1+\cos\theta_2 d\phi_2\Big)^2+
\frac{1}{6}r^2(d\theta_1^2+\sin^2\theta_1d\phi_1^2)\nonumber \\ &&
+\frac{1}{6}(r^2+6u^2)(d\theta_1^2+\sin^2\theta_2d\phi_2^2)\ ,
\end{eqnarray}
where
\begin{equation}
K=\frac{r^2+9u^2}{r^2+6u^2}\ .
\end{equation}
The $N$ $D3$-branes sourcing the warp factor are
located at the north pole of the finite $S^2$, i.e. at $r=0, \theta_2=0$. 
The corresponding warp factor  
$h$ is a function of both $r$ and $\theta_2$, which can be written as \cite{Klebanov:2007us}
\begin{equation} \label{KMwarp}
h=
L^4\sum_{l=0}^{\infty}(2l+1)H_l(r)P_l(\cos\theta_2)\;
\end{equation}
where $L^4={27\pi\over 4} g_s N(\alpha')^2$, 
$P_l(\cos\theta)$ is the $l$-th Legendre polynomial, and
\begin{equation}
H_l=\frac{2C_{\beta}}{9 u^2 r^{2+2\beta}}\quad_2F_1\Big(\beta,1+\beta,1+2\beta;-\frac{9u^2}{r^2}\Big)\ ,
\end{equation}
with the coefficients $C_{\beta}$ and $\beta$ given by
\begin{equation}C_{\beta}=\frac{(3u)^{2\beta}\Gamma(1+\beta)^2}{\Gamma(1+2\beta)}\ ,\qquad \beta=\sqrt{1+\frac{3}{2}l(l+1)}\ .\end{equation}
Using the Euclidean $D3$-brane located at fixed $\theta_2, \phi_2$,
it is possible to compute the VEV of the baryonic operator ${\rm det} B_2$.
Its modulus is found to be $\sim u^{\frac{3N}{4}}$ \cite{Klebanov:2007us}. 
Its phase, which is determined by value of the Goldstone boson of the 
broken $U(1)_B$ symmetry, will be discussed in section 5.2.

Far in the IR the gauge theory flows to the $\mathcal{N}=4$ SYM theory, as evidenced by the appearance
of an $AdS_5\times S^5$ throat near the location of the stack of the $D3$-branes. 
We will see that the gauge theory also contains an interesting additional sector coupled to 
this infrared CFT. The coupling of such an extra sector and an infrared CFT
is reminiscent of the unparticle physics scenarios \cite{Georgi:2007ek}.

\section{Global Strings}

In addition to the existence of a Goldstone boson, a hallmark of a broken $U(1)$ symmetry is the 
appearance of ``global'' strings, around which the Goldstone boson carries a non-trivial monodromy. 
We will show that on  
the supergravity side of the duality these global strings are partially wrapped $D3$-branes. 

Let us consider the IR (small $r$) region. There the warp factor
will approach some function $g^2$ whilst the resolved conifold contains a 2-sphere parametrized by $\theta_2, \phi_2$:
\begin{equation}
ds^2\rightarrow g^{-1}dx_{1,3}^2+gu^2d\Omega_2\ .
\end{equation}
Let us now consider a $D3$-brane whose world-volume spans the $t,x,\Omega_2$ coordinates, such that the brane wraps the 
two-sphere but remains extended along one of the $x$ directions. 
From the point of view of the dual gauge theory this will correspond to a string-like object 
along ${t,x}$. It is natural to identify it with the string originating from 
the breaking of $U(1)_B$, since its existence is connected 
with the finite $S^2$ at the bottom of the resolved conifold, 
which in turn requires that the $U(1)_B$ is broken. 

It is straightforward to compute the tension of such a wrapped $D3$-brane\footnote{
We set $\alpha'=1$ throughout the paper.}
\begin{equation}
\label{tension}
T_s=4\pi T_3u^2=\frac{u^2}{2\pi^2 g_s}\ .
\end{equation}
Interestingly, the tension is completely independent of the warp factor and 
remains finite at the bottom of the conifold.\footnote{
In section 5 we will show that, due to backreaction, the tension of
the string receives an additional logarithmically divergent contribution. This divergence is 
typical for an isolated co-dimension 2 global string.}
The global strings and the Goldstone modes they couple to belong
to another sector of the theory which ``sees'' the whole $S^2$, 
and is coupled to the $\mathcal{N}=4$ $SYM$. 
The very existence of this extra sector coupled to the CFT is an
interesting fact. It is reminiscent of the unparticle physics
scenarios \cite{Georgi:2007ek} typically 
characterized by a
a particle sector coupled to a conformal (unparticle) sector. 
Our construction amounts to a UV completion of such an unparticle scenario in terms of
the $SU(N)\times SU(N)$ SCFT with a VEV for one of the bi-fundamental fields.

\subsection{Kappa-symmetry for the $D3$}

The kappa-symmetry projection which our $D3$-brane should satisfy is

\begin{equation}
\Gamma_{\kappa}\epsilon=i\mathcal{L}_{DBI}^{-1}\gamma_4 \epsilon\ ,
\end{equation}

\noindent where $\gamma_4$ is the pull-back of the target space gamma matrices to the worldvolume of the brane, and $\epsilon$ is the background Killing spinor. 
Those spinors can be written in terms of the Killing spinor in the resolved conifold. For that, write the resolved conifold metric in terms of

\begin{center}
\begin{tabular}{c c}
$e^1=\frac{r}{\sqrt{6}}d\theta_1\ ,$ & $e^3=\frac{\sqrt{r^2+6u^2}}{\sqrt{6}}d\theta_2\ ,$\\
$e^2=\frac{r}{\sqrt{6}}\sin\theta_1d\phi_1\ ,$ & $e^4=\frac{\sqrt{r^2+6u^2}}{\sqrt{6}}\sin\theta_2d\phi_2\ ,$\\
$ e^5=\frac{\sqrt{K}r}{3}(d\psi+\cos\theta_1d\phi_1+\cos\theta_2d\phi_2)\ ,$ & $e^0=K^{-\frac{1}{2}}dr\ ;$
\end{tabular}
\end{center}

\noindent so that the 6d metric reduces to

\begin{equation}
ds_6^2=(e^0)^2+(e^1)^2+(e^2)^2+(e^3)^2+(e^4)^2+(e^5)^2\ ,
\end{equation}

It is then straightforward to see that for our brane wrapping $t,x,\theta_2,\phi_2$ and sitting at $r=0$, the kappa-symmetry projector above reduces to

\begin{equation}
\Gamma_{\kappa}\epsilon=i\Gamma_{tx}\Gamma_{34}\epsilon\ ,
\end{equation}
where $\Gamma_i$ are the flat space gamma matrices.

The background Killing spinor satisfies $\Gamma_{34}\epsilon=\Gamma_{12}\epsilon$ \cite{Cvetic:2000mh}
(for a review see \cite{Arean:2006nc}).
Since $i\Gamma_{34}$ and $i\Gamma_{12}$ are both projectors (they square to the identity and are traceless), these two equations can be solved to give

\begin{center}
\begin{tabular}{c c}
$i\Gamma_{34}\epsilon=\epsilon\ ,$ & $i\Gamma_{12}\epsilon=\epsilon\ ,$
\end{tabular}
\end{center}

\noindent so in the case at hand the kappa-symmetry projection reduces to

\begin{equation}
\Gamma_{\kappa}\epsilon=\Gamma_{tx}\epsilon=\epsilon\ .
\end{equation}

\noindent Now since $\Gamma_{tx}$ is a traceless matrix which squares to one, this condition is satisfied for half of the background spinors, so our strings are one-half BPS.

\section{The Fluctuation Containing the Goldstone Mode}

In order to gain more understanding of 
the strings we have found, we can consider the linearized backreaction 
in the background caused by our probe $D3$. Such a brane will source, to linearized order, 
a fluctuation in the 4-form RR potential containing the term
$a_2(x)\wedge W$ 
where $a_2$ is a 2-form in $R^{3,1}$ and $W$ is a closed 2-form 
in the resolved conifold: $dW=0$. 
This perturbation has to satisfy the linearized equations of motion, 
which read $d \delta F^{(5)}=0$ and $\delta F^{(5)}=\star \delta F^{(5)}$. 
We can ensure the latter by taking
\begin{equation}
\delta F^{(5)}=(1+\star)d (a_2(x)\wedge W)\ .
\end{equation}
Then the equations of motion reduce to 
\begin{equation}\label{eoma_2}
d\star_4da_2=0\ ,
\end{equation}
provided $W$ satisfies
\begin{equation}
d(h\star_6W)=0\ ,
\end{equation}
where $\star_4$, $\star_6$ are the Hodge duals with respect to the 
unwarped Minkowski and resolved conifold metrics, respectively.
Introducing the field $p(x)$ through $\star_4 da_2=dp$,  
we note that the fluctuation in the 5-form field strength reads 
\begin{equation}
\label{RRpert}
\delta F^{(5)}=da_2\wedge W+dp\wedge h\star_6W\ .
\end{equation}
The corresponding fluctuation of the 4-form potential is
\begin{equation}
\label{C4}
\delta C^{(4)}=
a_2(x)\wedge W + p\, h\star_6W\ . 
\end{equation}
We will see later on that the field $p(x)$ is the Goldstone boson for the broken $U(1)_B$.

\subsection{Equations for $W$}

Different types of forms in conifolds have been discussed in the literature 
(see for example \cite{Cvetic:2000mh}, \cite{Cvetic:2000db},  \cite{Lu:2002rk}). 
In our case, we are searching for
a closed 2-form which is co-closed upon multiplication with the warp factor. 
We will consider an ansatz that satisfies $dW=0$: 
\begin{equation}
W=\sin\theta_2 d\theta_2\wedge d\phi_2+d(f_1g^5+f_2\sin\theta_2d\varphi_2)\ ,
\end{equation}
where $f_1, f_2$ are functions of $r,\theta_2$.
It is convenient to define a re-scaled set of vielbeins 
\begin{center}
\begin{tabular}{c c}
$\epsilon^1_1=d\theta_1\ ,$ & $\epsilon^1_2=d\theta_2\ ,$\\
$\epsilon^2_1=\sin\theta_1d\phi_1\ ,$ & $\epsilon^2_2=\sin\theta_2d\phi_2\ ,$
\end{tabular}
\end{center}
and
\begin{equation}
g^5=d\psi+\cos\theta_1d\phi_1+\cos\theta_2d\phi_2\ .
\end{equation}
Then the 6-dimensional metric on the resolved conifold becomes
\begin{equation}
ds_6^2=K^{-1}dr^2+\frac{Kr^2}{9}(g^5)^2+
\frac{r^2}{6}(\epsilon^1_1)^2+\frac{r^2}{6}(\epsilon^2_1)^2+
\frac{r^2+6u^2}{6}(\epsilon^1_2)^2+\frac{r^2+6u^2}{6}(\epsilon^2_2)^2\ .
\end{equation}
After some algebra, the condition $d(h\star_6W)=0$ translates 
into two coupled differential equations for $f_1$ and $f_2$:
\begin{eqnarray}
\label{A}
&&\partial_r\Big(\frac{h\partial_rf_1 r(r^2+6u^2)}{12}\Big)+\frac{h}{3r(r^2+6u^2)}\Big(r^4(1-f_1+\frac{\partial_{\theta_2}(f_2\sin\theta_2)}{\sin\theta_2})-f_1(r^2+6u^2)^2\big)\nonumber\\ &+&\frac{r}{2K}\Big(\frac{1}{\sin\theta_2}\partial_{\theta_2}\big[h\sin\theta_2\partial_{\theta_2}f_1\big]\Big)=0;
\end{eqnarray}
and
\begin{equation}
\label{B}
\partial_{\theta_2}\Big[\frac{hr^3}{3(r^2+6u^2)}\Big(1-f_1+\frac{\partial_{\theta_2}(f_2\sin\theta_2)}{\sin\theta_2}\Big)\Big]+\partial_r\Big[\frac{hKr^3}{18}\partial_rf_2\Big]=0\ .
\end{equation}
These equations must be supplemented by a set of boundary conditions. 
Since the $\delta C^{(4)}$ fluctuation couples to the D3-brane  wrapped over
the $S^2$ at $r=0$, $W$ should approach there the volume form of the finite $S^2$. 
For that we require both $f_1(r=0,\theta_2), f_2(r=0,\theta_2)$ to vanish. On the other hand, as we will see later on, in the 
large $r$ region $W$ should asymptote to the usual $\omega_2$ 2-form in the singular conifold, where $f_1(r\rightarrow \infty,\theta_2)=\frac{1}{2}$ 
while $f_2(r\rightarrow \infty,\theta_2)=0$.

Interestingly equations (\ref{A})-(\ref{B}) come from minimizing the following functional
\begin{eqnarray}\label{I}
I&=&\int W\wedge h\star_6W=\int_0^\pi d\theta_2 \int_0^\infty dr \sin\theta_2\Big\{\frac{hr^3}{3(r^2+6u^2)}
\Big(1-f_1+\frac{\partial_{\theta_2}\big(f_2\sin\theta_2\big)}{\sin\theta_2}\Big)^2\nonumber\\ &+&\frac{hr}{12}(r^2+6u^2)\big(\partial_r f_1\big)^2+\frac{hr}{2K}\big(\partial_{\theta_2}f_1\big)^2+\frac{hKr^3}{18}\big(\partial_rf_2\big)^2+\frac{h}{3r}(r^2+6u^2)f_1^2\Big\}\ .
\end{eqnarray}
With the boundary conditions above, one can check that $I$ remains finite using the warp factor in 
\cite{Klebanov:2007us}. 
For small $r$ the leading term in the warp factor, $h\sim {1\over r^2}\delta (1-\cos \theta_2)$, can
be used to show that no divergence occurs. 
Had we assumed that $f_1$ was non-zero at the tip then $I$ would not converge due to the last term. 
The same argument holds for $f_2$, since otherwise $I$ would diverge at $\theta_2=\pi$. 

For large $r$ the leading behavior of the warp factor, $h=L^4/r^4$, 
renders the integral convergent provided $f_1$ approaches a constant 
(otherwise the term with $\partial_rf_1$ would blow up), while (\ref{A}) sets it to $1/2$
(in the unwarped case, the integral would instead diverge).

We have not been able to find an analytic solution for (\ref{A})-(\ref{B}). However since with the chosen boundary conditions $I$ remains finite, 
we expect that such a solution exists and is unique. 
It could be looked for using numerical relaxation algorithms applied to the functional $I$. 
We now turn to the analysis of (\ref{A})-(\ref{B}) in several limits and check that they give sensible results.

\subsection{Unwarped resolved conifold}

If we set the warp factor to $h=1$, then the equations we are solving describe the
harmonic 2-form $W_{\rm harm}$ on the resolved conifold. 
This problem, in a slightly different context, has already been discussed in \cite{Lu:2002rk} 
with similar results. Note that in the case of $h=1$, (\ref{I}) is no longer convergent and therefore the solution for W is not normalizable.

In this instance we can take $f_1, f_2$ to depend only on $r$. Then (\ref{B}) shows that $f_2$ may be set to
zero, so we are left with a simplified version of (\ref{A}) which reads 

\begin{equation}
\partial_r\Big(\frac{\partial_rf_1 r(r^2+6u^2)}{12}\Big)+\frac{1}{3r(r^2+6u^2)}\Big(r^4(1-f_1)-
f_1(r^2+6u^2)^2\Big)=0\ .
\end{equation}
The solution to this equation is \cite{Lu:2002rk} 
\begin{equation}
f_1=\frac{r^2}{2(r^2+6u^2)}\ ,
\end{equation}
giving
\begin{equation}
W_{\rm harm} = \left ({ 1\over 2} +{3u^2\over r^2+ 6 u^2}\right )\sin\theta_2 d\theta_2\wedge d\phi_2
-\left ({ 1\over 2} -{3u^2\over r^2+ 6 u^2}\right )\sin\theta_1 d\theta_1\wedge d\phi_1
+{6 u^2 r\over (r^2+ 6 u^2)^2} dr\wedge g_5
\ .
\end{equation}
In the UV (at large $r$), this form approaches
the harmonic 2-form on the singular conifold, 
\begin{equation}
\omega_2=\frac{1}{2}(\sin\theta_2d\theta_2\wedge d\phi_2-\sin\theta_1d\theta_1\wedge d\phi_1)\ .
\end{equation}
Thus the harmonic form $W_{\rm harm}$ on the resolved conifold interpolates between
$\sin\theta_2 d\theta_2\wedge d\phi_2$,
which is the volume form of the $S^2$ at $r=0$,
 and $\omega_2$ at large $r$.

We note that, even in the cases where the warp factor is non-trivial, $W_{harm}$ can
be used to construct the following solutions of the equations of motion:
\begin{equation}
B_2=\theta(x) W_{\rm harm}\ ,
\end{equation}
where $\theta(x)$ is a function of the Minkowski coordinates 
only. An analogous solution also exists for
$C_2$.
In the equation of motion 
\begin{equation}
d\star dB_2=0 
\end{equation}
the warp factor cancels, and it reduces to
to $d\star_4d\theta=0$.
As anticipated one can check that this mode of $B_2$ is not normalizable. This means that it corresponds to a change in the 
Lagrangian of the dual gauge theory. Indeed choosing a constant $\theta$ corresponds to
changing $g_1^{-2}- g_2^{-2}$, where $g_i$ are the gauge couplings of the $SU(N)\times SU(N)$
gauge theory \cite{Klebanov:1998hh}.

We have not been able to find an analytic solution 
to (\ref{A}) and (\ref{B}) with the warp factor (\ref{KMwarp}). 
However we can analyze the behavior of the solutions in the asymptotic regimes. 

\subsection{IR limit: $r\rightarrow 0$}

The boundary conditions in the IR are such that $W$ approaches
the volume form of the finite 2-sphere. This requires that both $f_1$ and $f_2$ vanish at the tip of the cone. 
In the small $r$ limit, (\ref{A}) reads
\begin{equation}
\label{AIR}
\partial_r\Big(\frac{hru^2\partial_rf_1 }{2}\Big)+\frac{h}{18u^2r}\big(r^4-36u^4f_1\big)=0\ .
\end{equation}
In turn, for equation (\ref{B}) we have
\begin{equation}
\label{BIR}
\partial_{\theta_2}\Big[hr^3\Big]+\partial_r\Big[ \frac{9u^2}{6}hr^3\partial_rf_2\Big]=0\ .
\end{equation}
In order to proceed further, we need the explicit form of the warp factor in the IR. 
It is known from \cite{Klebanov:2007us} that its behavior depends crucially on the point 
in the sphere we are considering. As argued in \cite{Klebanov:2007us}, for a small
distance $y$ from the north pole,
$h\sim y^{-4}$.  
For small $r$ and $\theta_2$, $y^2\sim 2r^2/3+u^2\theta_2^2$, and thus
\begin{equation}
\label{hir}
h\sim \frac{9 L^4}{4 (r^2+3u^2\theta_2^2/2)^2}\ .
\end{equation}
  
In order to analyze the behavior of the fluctuation near the north pole, 
we will start by  
setting $\theta_2=0$ and keeping a very small $r$. In this case from (\ref{hir})
$\partial_{\theta_2}h=0$. Upon taking $h=\frac{9 L^4}{4r^4}$, (\ref{BIR}) becomes
\begin{equation}
\partial_r\Big(\frac{\partial_rf_2}{r}\Big)=0\ .
\end{equation}
The appropriate solution is $f_2\sim r^2$ which indeed vanishes when $r$ approaches zero.
We turn now to equation (\ref{AIR}) for $\theta_2=0$. It is easy to show that $f_1=\frac{r^4}{36u^4}$. Note that $f_1$ also goes to zero with $r$. 

It is instructive to consider an alternative way of reaching the north pole, namely running along 
the $S^2$ at $r=0$ towards $\theta_2=0$. 
From (\ref{AIR}) we see that setting $r=0$ requires us to set $f_1=0$ in the whole $S^2$. Also at $r=0$
and small $\theta_2$, the warp factor in (\ref{hir}) leads to $h= L^4(u\theta_2)^{-4}$. 
Then equation (\ref{BIR}) reduces to
\begin{equation}
\frac{1}{u^2\theta_2^4}\partial_r(r^3\partial_rf_2)=0\ .
\end{equation}
Therefore $\partial_r(r^3\partial_rf_2)=0$ at $r=0$, which sets $f_2$ to be constant which we choose to vanish.

\subsection{The UV limit: $r\rightarrow\infty$}

For large $r$ the warp factor approaches that of the singular conifold, namely $h=\frac{L^4}{r^4}$. However in this case the subleading corrections can be written as a series expansion in powers of $1/r$, so it is possible to truncate the series at a certain $l$. Let us consider not just the pure UV behavior but also the first correction, which already exhibits angular dependence
\begin{equation}
h=\frac{L^4}{r^4}+\frac{9u^2L^4\cos\theta_2}{r^6}+ \ldots \ .
\end{equation}
Denoting $f_1$ and $f_2$ as the relevant functions in the large $r$ region, it is straightforward to see that they should
be of the form
\begin{equation}
f_1=\frac{1}{2}+F_1(\theta_2,r)\ ,\qquad f_2=F_2(\theta_2,r)\ ;
\end{equation}
where $F_i$ are terms which vanish in the asymptotic limit. 
Therefore both $f_1$ and $f_2$ satisfy the required boundary conditions in the $UV$. 
The equations (\ref{A}), (\ref{B}) read
\begin{equation}
\label{AUV}
\partial_r\Big[\frac{\partial_rF_1}{12r}\Big]+\frac{1}{3r^3}\Big[-2F_1+\frac{3}{2}
\frac{\partial_{\theta_2}\big(\sin\theta_2\partial_{\theta_2}F_1\big)}{\sin\theta_2}
+\frac{\partial_{\theta_2}\big(F_2\sin\theta_2\big)}{\sin\theta_2}\Big]-\frac{2u^2}{r^5}=0\ ;
\end{equation}
and
\begin{equation}
\label{BUV}
\partial_r\Big[\frac{\partial_rF_2}{6r}\Big]+\partial_{\theta_2}\Big[\Big(\frac{1}{r^3}+\frac{\big(9u^2\cos\theta_2-6u^2\big)}{r^5}\Big)\Big(\frac{1}{2}-F_1+\frac{\partial_{\theta_2}\big(F_2\sin\theta_2\big)}{\sin\theta_2}\Big)\Big]=0\ .
\end{equation}
The form of these equations suggests that
\begin{equation}
F_1=\frac{u^2}{r^2}A(\theta_2)\ ,\qquad F_2=\frac{u^2}{r^2}B(\theta_2)\ .
\end{equation}
Then  (\ref{AUV}) reduces to
\begin{equation}
\frac{\partial_{\theta_2}\big(\sin\theta_2\partial_{\theta_2}A\big)}{\sin\theta_2}+\frac{2}{3}\frac{\partial_{\theta_2}\big(B\sin\theta_2\big)}{\sin\theta_2}-4=0\ ,
\end{equation}
which can be easily integrated to give  the first order equation
\begin{equation}
\label{angB}
\frac{1}{2}\sin\theta_2\partial_{\theta_2}A+\frac{B}{3}\sin\theta_2=-2\cos\theta_2+k\ ,
\end{equation}
where $k$ is a constant of integration. 
We can now plug this into (\ref{BUV}), and get a single differential equation (which is third order) for $A$
\begin{equation}
\label{angA}
\frac{4k}{\sin\theta_2}-\frac{8}{3}\frac{\cos\theta_2}{\sin\theta_2}-\frac{9}{2}\sin\theta_2-\frac{5}{3}\partial_{\theta_2}A-\frac{3}{2}\partial_{\theta_2}\Big[\frac{\partial_{\theta_2}\big(\sin\theta_2\partial_{\theta_2}A\big)}{\sin\theta_2}\Big]=0\ .
\end{equation}
Let us mention that it is possible to write down the next order terms, which will still depend on rational powers of $r$. 
It suggest that the next correction is going to go like $\frac{1}{r^4}$. However at the next following order, we will already encounter irrational powers of $r$ coming from the fact that $h$ contains them. 
(\ref{angA}) is most easily solved in terms of $F=\sin\theta_2\partial_{\theta_2}A$ through the equation
\begin{eqnarray}
\label{F}
\sin\theta_2\partial_{\theta_2}A&=&-\tilde{A}\, _2F_1\left(-\frac{5}{6},\frac{1}{3},\frac{1}{2};\cos^2\theta_2\right)\\ \nonumber &-&\tilde{B}\, _2F_1\left(-\frac{1}{3},\frac{5}{6},\frac{3}{2};\cos^2\theta\right)\ \cos\theta_2+\frac{27}{8}\sin^2\theta_2-\frac{8}{5}\cos\theta_2+\frac{12k}{5}\ .
\end{eqnarray}
where now $\tilde{A}$ and $\tilde{B}$ are more integration constants. Using (\ref{angA}) in (\ref{angB}) gives $B\sin\theta_2$.

For reasons that will be clarified later we would like to keep $\partial_{\theta_2}(B\sin\theta_2)$ finite, which in turn requires that $\partial_{\theta_2}A$ is 
also finite. Now given (\ref{F}), in order to for this to hold we have to impose boundary conditions such that the r.h.s vanishes
at both $0$ and $\pi$. This fixes

\begin{equation}
\label{BC}
\tilde{A}=\frac{2 k}{5\sqrt{\pi}}\Gamma\left(\frac{1}{6}\right)\,\Gamma\left(\frac{3}{4}\right)\ , \qquad \tilde{B}=\frac{24}{5\sqrt{\pi}}\,\Gamma\left(\frac{2}{3}\right)\,\Gamma\left(\frac{11}{6}\right)\ .
\end{equation}
Let us point out that in the asymptotic UV region we find $W=\omega_2$, exactly as in the unwarped toy model.
\section{Spontaneous Breaking of the Baryonic Symmetry} 

In a field theory with spontaneously broken $U(1)$ symmetry, the 
classical value of the $U(1)$ current is
\begin{equation} \label{currentev}
J_{\mu}^{cl}\sim \frac{i}{2}\big(\Phi^*\partial_{\mu}\Phi-\Phi\partial_{\mu}\Phi^*\big)
= |\Phi_0|^2\partial_{\mu}\pi(x)\ ,
\end{equation}
where we substituted $\Phi=\Phi_0 e^{i\pi(x)}$, and $\pi(x)$ is the Goldstone field. Let us show how this expectation value appears for
the $U(1)$ baryonic current using the AdS/CFT correspondence. 
At large $r$ the perturbation $\delta F_5$ behaves as
\begin{equation} \label{asymp}
\delta F^{(5)}\rightarrow (1+\star) r^{-3} dr\wedge dp\wedge \omega_3\ .
\end{equation}
Therefore the leading term in $\delta C^{(4)}$ contains $r^{-2} dp \wedge \omega_3$ 
(note that this corresponds to a different gauge choice from that in (\ref{C4})).
We also know that the massless gauge field $A^B_\mu$ dual to the baryonic current 
enters as $\delta C^{(4)}= A^B\wedge \omega_3$. 

It follows that 
\begin{equation}
A^B_\mu(r)\rightarrow  r^{-2} \partial_\mu p\ .
\end{equation}
This is a normalizable perturbation near the boundary of AdS$_5$ that, through the
AdS/CFT dictionary \cite{Klebanov:1999tb}, 
implies a relation of the form (\ref{currentev}).\footnote{
Essentially the same argument applies to the breaking of the
baryonic symmetry in the cascading gauge theory. Using the perturbation containing the
Goldstone mode found in \cite{Gubser:2004qj} we observe the same asymptotic behavior as in
(\ref{asymp}) up to powers of $\ln r$ characteristic of the cascading theory.
This again implies the expectation value of the baryonic current (\ref{currentev}).}

As for any Goldstone boson, it is interesting to determine its ``decay constant'' $f_p$ which
appears in the 4-d effective action as
\begin{equation}
\label{4dSeff}
f_p^2 \int d^4 x dp\wedge \star_4 dp
\end{equation}
In deriving this action from dimensional reduction of the type IIB action, we face the usual
problem that $\delta F^{(5)}$ containing the Goldstone boson perturbation is self-dual so that
$\delta F^{(5)}\wedge \star \delta F^{(5)}=0$. Instead we will adopt the approach of
removing the self-duality constraint and using 
$\delta F^{(5)}=dp\wedge h\star_6W$. Now the action no longer vanishes, and we find
\begin{equation}
f_p^2 \sim {1\over g_s^2 } \int h W\wedge \star_6 W
\end{equation}
Note that from (\ref{I}), $f_p^2\sim g_s^{-2} I$. Therefore, in the warped case, 
$f_p^2$ is a finite quantity determined by the minimum of the functional I.
With the analysis of the leading asymptotic corrections we can examine the UV behavior of 
the integrand in $I$
\begin{equation}
\label{W^2}
\int dr \sin\theta_2\, \Big(\frac{L^4}{6r^3}+\frac{L^4u^2}{6r^5}\Big(9\cos\theta_2+2\frac{\partial_{\theta_2}\big(B\sin\theta_2\big)}
{\sin\theta_2}\Big)+\mathcal{O}(r^{-7})\Big)
\end{equation}
Note that with the choice of boundary conditions (\ref{BC}), we find
\begin{equation}
\label{Bsin}
\partial_{\theta_2}\Big(B\sin\theta_2\Big)|_{\theta=0,\pi}=0\ .
\end{equation}
Therefore the first correction of order $\mathcal{O}(r^{-5})$ in (\ref{W^2}) vanishes upon integration. Note also that (\ref{BC}) also renders $\partial_{\theta_2}A$ finite through the relationship in (\ref{F}). Indeed with this choice of boundary conditions, the $\delta F_5$ remains under 
control on the whole $S^2$.

For scales larger than the resolution length $u$, we expect the geometry to approach that of the singular conifold. In turn, for scales smaller than $u$, we expect the resolution of the geometry 
to take over and smoothly close the cone.  Thus, in order to estimate the decay constant, 
we will take the asymptotic value (\ref{W^2}) and cut off the radial integral at $r\sim u$.  
We find the decay constant goes like
\begin{equation}
f_p^2 \sim {L^4 \over u^2 g_s^2 }\sim {N \over g_s u^2 }
\end{equation}
Therefore $f_p$ blows up in the limit where $u$ vanishes. 
Using this we can define a normalized Goldstone boson field $\tilde p=p f_p$, in terms of which
the VEV of the current takes the canonical form  
$\langle J^B_\mu \rangle\sim f_p \partial_\mu \tilde p $, in agreement with (\ref{currentev}).
\subsection{Global strings from baryonic symmetry breaking}

As argued in section 3, a $D3$-brane wrapped over the 2-cycle of the resolved conifold 
is dual to the global string in the gauge theory
which arises due to the baryonic symmetry breaking. Now that we have found
the supergravity fluctuation (\ref{C4}) of the 4-form gauge potential, which contains the
Goldstone boson, we can provide further support for this identification.

The string is charged under the 2-form potential $a_2$ in Minkowski space. 
Writing the Minkowski metric as
\begin{equation}
dx^2_{1,3}=-dt^2+dx^2+d\rho^2+\rho^2d\theta^2\ ,
\end{equation}
we see that a string extended along $t,x$ and localized at
$\rho=0$ produces $a_2=a(\rho)dt\wedge dx$. The equation of motion following from the 4-dimensional effective action (\ref{4dSeff}) can be recast in terms of $a$, and reads
\begin{equation}
-\partial_{\rho}(\rho\partial_{\rho}a)\sim
\frac{T_s}{f_p^2} n \delta(\rho) \ ,
\end{equation}
where $n$ is the integer winding number of $D3$ around the $S^2$. 
Here $T_s$ stands for the 4d tension of the string in (\ref{tension}). This equation can be integrated to give
\begin{equation}
a(\rho)\sim
\frac{T_s}{f_p^2}n\log\rho\ .
\end{equation}
We also note that $\star_4da_2=d(T_s n \theta/f_p^2)$, giving $p=T_s n \theta/f_p^2$. This
exhibits the expected monodromy of the Goldstone boson upon encircling the string. 
Furthermore we can compute the energy density 
due to the back-reaction of the string: 
\begin{equation}
\mathcal{E}=  f_p^2\int^{\rho_{\rm max}} 
(\partial_\mu p)^2 \rho d\rho d\theta\sim \frac{2\pi^2T_s^2}{f_p^2} n^2 \log\rho_{\rm max}\ .
\end{equation}
Thus we see that the energy density exhibits a logarithmic divergence, which is typical for a global string of codimension two. 
It is worth noting that $\mathcal{E}$ is of order $N^0$, while both $f_p^2$, $T_s$ are of order $N$.

\subsection{The phase of the baryonic condensate}

As shown in \cite{Klebanov:2007us} the expectation values of baryonic operators can be deduced
from the action of certain Euclidean $D3$-branes ($E3$-branes).
The particular warped resolved conifold studied in \cite{Klebanov:2007us}, corresponding to
all $D3$-branes placed at the north pole of the $S^2$ at $r=0$, 
is dual to the $SU(N)\times SU(N)$ gauge theory with a VEV for
a bi-fundamental field $B_2$. To calculate the VEV of
the gauge invariant baryonic operator ${\rm det}\ B_2$, we need to consider
$e^{-S_{E3}}=e^{-S_{DBI}}e^{-S_{CS}}$
for the $E3$-brane wrapping the 4-chain with internal coordinates $r,\psi,\theta_1,\phi_1$,
located at fixed $\theta_2,\phi_2$. 
The phase of $\langle {\rm det} B_2 \rangle $ comes from 
the factor $e^{-S_{CS}}$, since for a Euclidean embedding the Chern-Simons term is imaginary.
This fact was important also for computing baryon VEV's in the cascading gauge theory,
where they were related to Euclidean D5-branes \cite{Benna}. For the resolved conifold case,
where baryonic operators correspond to $D3$-branes,

\begin{equation}
S^{CS}=iT_3\int P[C^{(4)}]\ .
\end{equation}
We expect that the phase of the baryonic condensate is proportional to the
Goldstone boson field $p$.
Indeed, using the $\delta C^{(4)}$ (\ref{C4}) containing the Goldstone boson,
we find that only the second term contributes to the integral:
\begin{equation}
S^{CS}=i16\pi^2T_3 p\int_0^\infty 
dr \frac{hr^3}{3(r^2+6u^2)}\Big(1-f_1+\sin^{-1}\theta_2\partial_{\theta_2}(f_2\sin\theta_2)\Big)\ .
\end{equation}
In order to check the convergence of the integral, we note that in the UV $f_2$ vanishes and $f_1$ 
approaches $1/2$. The integrand approaches $L^4/(6 r^3)$, rendering the integral convergent. We can estimate the integral as before, cutting-off at $r=u$ and using the asymptotic values for $f_1, f_2$. We find
\begin{equation}
S^{CS}\sim i \frac{9 p N}{8 u^2}\ .
\end{equation}
Thus although we cannot perform the full integral, we see that the phase of the baryonic condensate 
behaves as we expected, namely it is proportional to the axion $p$. 

\section{Final comments}

In this note we studied breaking of the $U(1)$
baryonic symmetry in the warped resolved conifold, and the 
Goldstone boson it produces. 
This field might acquire some non-trivial monodromy in the 4-dimensional Minkowski space and 
generate a global string. 
We have found such a string and proposed the equations which the axionic Goldstone boson should satisfy. 
Our results depend on the existence of a very particular mode involving a 2-form in the resolved conifold. 
This 2-form, $W$, has to interpolate between the 
volume form of the finite $S^2$ at the bottom of the conifold, and the $\omega_2$ 2-form in the singular conifold to which the geometry asymptotes. 
Unfortunately, we have not been able to find this 2-form analytically in the warped case.
However, we have provided some evidence that, in the asymptotic regimes, the equations 
admit a solution which behaves as expected. 
In any case we have shown that the equations which this 2-form satisfies 
can be derived from minimizing a functional $I$. 
With the appropriate boundary conditions 
this functional remains finite, supporting the existence of such a 2-form. 
Additionally, since its norm is given by $I$, our fluctuation is a normalizable mode. 
The value of $I$ also controls the decay constant $f_p$ of the Goldstone boson. 
We note that $I$ is divergent for the unwarped solution and is finite only for the warped solution.

It would be interesting to find the full solution for $W$. 
Since we have reduced finding $f_1$ and $f_2$ to a variational problem, 
it should be possible to use a numerical relaxation method with the functional $I$.

It is worth stressing that the discussion in this note concerns the case of non-compact warped
conifold  dual to the gauge theory where 
the baryonic symmetry is
global. 
 It would be interesting to consider embedding this set-up into a full string compactification. 
In that case one would expect the baryonic symmetry to be gauged, and our strings to appear as local ones. 
The discussion goes in very much the same 
spirit as that in the case of a pure CY compactification \cite{Greene:1995hu, Greene:1996dh}. 
Since we are considering the resolved conifold, the 
would-be gauge $U(1)_B$ is in the Higgs phase. In turn this implies a linear potential for the monopole-antimonopole interaction. 
One would be tempted to identify the string connecting the monopoles with our string, 
which would then appear as a local string.\footnote{At the SUGRA level, similar strings to these have been discussed in for example \cite{tomas1, tomas2}.} 
However the local string is actually a combination of strings obtained by wrapping branes 
on a set of cycles that sum up to zero in homology. It would be interesting to 
adapt the analysis in \cite{Greene:1995hu, Greene:1996dh} to the case at hand, 
in which there is a non-trivial warp factor and 5-form flux. 
Also, the $\delta C^{(4)}$ mode
found here might be responsible for a higher-form mediation of SUSY breaking 
along the lines of \cite{Verlinde:2007qk}.
A key ingredient in this set-up is the existence of two homologous 2-cycles
allowing for a combination of two $U(1)$ gauge fields 
to remain massless. 
This combination appears in the low energy theory and can potentially mediate SUSY 
breaking between the hidden and visible sectors. Perhaps such a construction can be implemented
in a flux compactification with two 
different warped resolved conifold throats separated by some distance within the compact dimensions. 

A related motivation for our work is that        
the partially wrapped $D3$-branes may provide models for cosmic strings in warped compactifications 
with resolved conifold throats. In such models, the smallness of the string tension is not due to 
the warp factor, but is rather due to the smallness of the resolution parameter $u$ 
which also sets the scale of the string tension. 
Such a construction seems to provide
a realization of the unparticle physics
scenario \cite{Georgi:2007ek}, since after the RG flow to the IR, the resulting theory contains an interacting CFT (the $\mathcal{N}=4$ theory) coupled to a non-conformal sector whose scale is set by $u$.

It would also be interesting to extend the results obtained here to more general 
resolved CY cones, such as those discussed in \cite{Martelli}.

 \section*{Acknowledgments}

We would like to thank S. Benvenuti for collaboration at the initial stages of this project,
and C. Herzog for comments on the manuscript. We thank D.Malyshev and I.Yavin for useful discussions.
D. R-G would like to thank F. Marchesano, T. Ort\'{\i}n, A. V. Ramallo and H. Verlinde for enlightening discussions, and CERN Theory Division for warm hospitality while this work was being completed. J.W 
thanks S. Thomas, A. Ritz, L. Thorlacius and the University of Iceland for their hospitality 
whilst this work was completed. This research was supported
in part by the
National Science Foundation under Grant No. PHY-0243680. 
D. R-G. acknowledges financial support from the European Commission through Marie Curie OIF grant contract no. MOIF-CT-2006-38381.
Any opinions, findings, and conclusions or
recommendations expressed in this material are those of the
authors and do not necessarily reflect the views of the National
Science Foundation.

\end{document}